\documentclass[letterpaper]{aa}
\usepackage[dvips]{graphicx}
\usepackage{psfig}
\newcommand{\pas}{.\hskip-2pt$^{\prime\prime}$}

\def\FE{[Fe{\sc ii}]}

\begin{document}
\title{Discovery of [FeII]- and H$_2$-emission from protostellar jets in the 
CB3 and CB230 globules}
\author{F. Massi \inst{1} \and C. Codella \inst{2} \and J. Brand \inst{3}}
\institute{
Osservatorio Astrofisico di Arcetri, INAF, Largo E. Fermi 5,
I-50125 Firenze, Italy
\and
Istituto di Radioastronomia, CNR, Sezione di Firenze, Largo E. Fermi 5,
I-50125 Firenze, Italy
\and
Istituto di Radioastronomia, CNR, Sezione di Bologna, Via P. Gobetti 101,
I-40129 Bologna, Italy
 }
\offprints{F. Massi, \email{fmassi@arcetri.astro.it}}
\date{Received date; accepted date}

\titlerunning{Protostellar jets in CB3 and CB230}
\authorrunning{F. Massi et al.}

\abstract{
Four Bok globules were studied in the Near-Infrared, through narrow-band 
filters, centered at the 1.644~$\mu$m line of [FeII], the H$_2$-line 
at 2.122~$\mu$m, and the adjacent continuum. 
\hfill\break\noindent
We report the discovery of \FE\ and H$_2$ protostellar jets and knots 
in the globules CB3 and CB230. The [FeII]-jet in CB230 is defined by a
continuous elongated emission feature, superimposed on which two knots are 
seen; the brighter one lies at the tip of the jet. The jet is oriented in the
same direction as the large-scale CO outflow, and emerges from the nebulosity
in which a Young Stellar Object is embedded. The H$_2$ emission 
associated with this jet is fainter and wider than the [FeII] emission, and 
is likely coming from the walls of the jet-channel.
\hfill\break\noindent
In CB3 four H$_2$ emission knots are found, all towards the blue-shifted lobe
of the large-scale outflow. 
There is a good correspondence between the 
location of the knots and the 
blue-shifted SiO(5$-$4) emission, 
confirming that SiO emission is tracing the jet-like flow rather well.
\noindent
No line emission is found in the other two targets, CB188 and CB205, although
in CB205 faint line emission may have been hidden in the diffuse nebulosity
near the IRAS position. Around this position a small group of ($\geq 10$) 
stars is found, embedded in the nebula. A diffuse jet-like feature near this 
group, previously reported in the literature, has been resolved into 
individual stars.
\keywords{Stars: formation -- ISM: jets and outflows -- Infrared: ISM: lines 
and bands -- ISM: individual objects: CB3 -- ISM: individual objects: CB230 --
ISM: individual objects: CB188 -- ISM: individual objects: CB205}
}

\maketitle

\section{Introduction}
\label{int}
Outflow and infall are inextricably associated with the very earliest stages
of star formation. Even while still accreting matter, a newborn star generates
a fast, well-collimated stellar wind that forms jets which sweep up the
ambient molecular gas, creating bipolar molecular outflows. The
high-velocity winds create shocks, which heat the gas to thousands of K
when breaking up into the ambient gas.
Understanding the details of the mechanism producing the acceleration
of the outflow is fundamental to the understanding of the star formation
process itself.

\noindent
Emission of millimetre molecular lines (usually CO) at excitation temperatures
T$_{\rm ex} \approx 10-20$~K, is used to study the outflow's large-scale
morphology. Whereas this component consists mostly of swept-up cloud 
material, 
and as such offers a time-averaged picture, 
the {\it present} flow activity is represented by a fast, hot component,
traced by H$_2$ ro-vibrational lines at $T_{\rm ex} \sim 2000$~K. Where this
jet-like component interacts with the slower flowing gas or the ambient cloud,
bow-shaped shock fronts are visible. This component, at T$_{\rm ex} \sim
100$~K, can be traced by mm-emission of molecular species that are produced
only in a shock-driven chemistry (e.g Bachiller~\cite{bach}).

The link between the 10-100~K gas and the hot
jet-component is not well understood. In particular, it is still an open
question how the efficiency of the processes leading to chemical anomalies
depends on the shock-type (J/C) and -characteristics. The comparison between
the shocked gas components at different temperatures is thus fundamental to
the study of the jet/outflow system and the energetics of the star forming
process.

\begin{figure}
\resizebox{7.5cm}{!}{\rotatebox{270}{\includegraphics{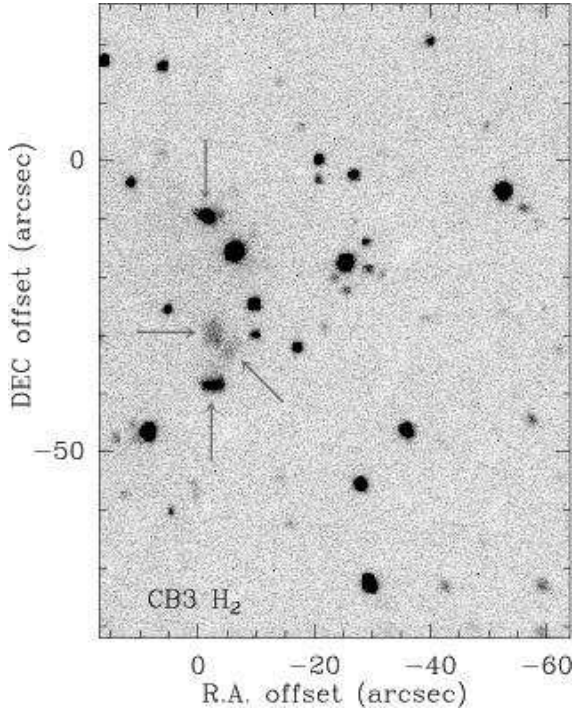}}}
\caption[]{TNG/NICS observations of CB3: The H$_2$ emission (still including 
the contribution from the continuum). The 4 knots (indicated by arrows) of
H$_2$ line-emission are not at all visible in the K-continuum image (not 
shown).
}
\label{cb3h2kcont}
\end{figure}

\begin{figure}
\resizebox{8.5cm}{!}{\rotatebox{270}{
\includegraphics{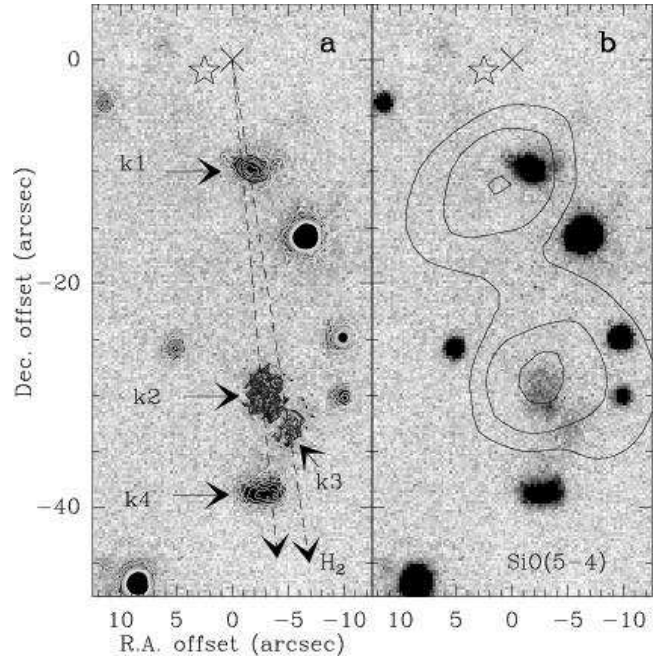}}}
\caption[]{{\bf a}\ A zoom-in on the four H$_2$ knots
in CB3. To better outline k2 and k3, contours for them are drawn with smaller 
increments than for the other knots in the panel; lowest contour is at 
$\sim 2.5\sigma$.
The dashed lines indicate the alignment of pairs of knots with the outflow
centre. The centre of the outflow (Codella \&
Bachiller~\cite{codbach}) lies at offset (0,0) and is indicated by a cross.
The 1.3-mm peak (Launhardt \& Henning~\cite{launhen}) is indicated by the star.
{\bf b}\ As a, but overlaid with the contours of integrated SiO(5--4)
emission at $-42.5 (\pm 0.5)$~km s$^{-1}$
(Codella \& Bachiller~\cite{codbach}). The angular resolution of the SiO map 
is 11$\arcsec$.
}
\label{cb3h2}
\end{figure}

\smallskip\noindent
A very good place to study the jet-component, the large-scale outflow, and
their interaction are the Bok globules: cold
(10 K) and
relatively isolated molecular clouds associated with star formation.
A catalogue of such objects (at $\delta > -30^{\circ}$) was compiled by
Clemens \& Barvainis (\cite{clemens}; hereafter CB).
Because of relatively simple structure, globules form mainly low-mass stars 
in small numbers, and are therefore without the observational confusion that 
one encounters in regions like Orion or Ophiuchus. 

With the italian TNG (Telescopio Nazionale Galileo) we have searched in four 
globules 
for the jet-component in the Near-InfraRed (NIR) through narrow-band H$_2$ 
(2.122~$\mu$m) and [FeII] (1.644~$\mu$m) filters.
These two lines are particularly useful, as [FeII] traces
(J-) shocks with velocities of a few 100~km s$^{-1}$ and is therefore expected
to outline the inner jet-channel, closest to the driving source of the flow.
Molecular hydrogen, which dissociates at shock velocities 
$>25-45$~km\,s$^{-1}$, traces
slower (C-) shocks and is excited in bow shocks and in shock-wakes (e.g. Allen
\& Burton~\cite{allen}) and is therefore a good probe for the region of 
interaction between jet and ambient material.
All four globules have been observed in the broad-band NIR (Yun \& 
Clemens~\cite{yc94a}) and detected at 1.3~mm continuum
(Launhardt \&
Henning~\cite{launhen}) and were found to contain embedded objects.
Molecular outflows were also found in all four objects (Yun \& 
Clemens~\cite{yc94b}; Codella \& Bachiller~\cite{codbach}). Together, these 
findings identify these globules 
as good candidates in a search for protostellar jets. 
In this paper we report the detection of H2 and [FeII] line emission knots
and jets in two of the four globules. 

\section{Observations and data reduction}
\label{obs}

Observations were carried out with the 3.58-m Italian Telescopio Nazionale 
Galileo (TNG) at La Palma (Canary Islands, Spain) on July 13-14, 2002.
The images were obtained with the Near Infrared Camera Spectrometer 
(NICS; Baffa et al. \cite{baffa}) 
through narrow-band filters centered on the 1.644~$\mu$m \FE\ 
$a ^4D_{7/2} - a ^4F_{9/2}$ and 2.122~$\mu$m H$_2$ v=$1-0$ S(1) lines and on 
the adjacent continuum at 1.57 $\mu$m (H$_{\rm cont}$) and 2.28 $\mu$m 
(K$_{\rm cont}$).
NICS is based on a 1024 $\times$ 1024 HgCdTe Hawaii array detector; 
we used a scale of 0$\farcs$25~pixel$^{-1}$, resulting in a field-of-view of 
$4\farcm2 \times 4\farcm2$.
The 4 fields to be imaged were centered at the coordinates listed in
Table~\ref{coords}, taken from Yun \& Clemens (\cite{yc92}, \cite{yc94b}) and 
Codella \& Bachiller (\cite{codbach}) in order to cover all or most of the 
area of the molecular outflows.

\begin{table}
\caption[]{Centered coordinates of the observed globules}
\label{coords}
\begin{flushleft}
\begin{tabular}{llllrll}\hline
\multicolumn{1}{l}{Globule} & \multicolumn{1}{l}{} &
\multicolumn{1}{c}{$\alpha$} & \multicolumn{2}{c}{(J2000)} &
\multicolumn{1}{c}{$\delta$} & \multicolumn{1}{l}{} \\
\multicolumn{1}{l}{} & \multicolumn{1}{c}{{\sl h}} & 
\multicolumn{1}{c}{{\sl m}} & \multicolumn{1}{c}{{\sl s}} &
\multicolumn{1}{c}{$\degr$} & \multicolumn{1}{c}{$\arcmin$} &
\multicolumn{1}{c}{$\arcsec$} \\
\hline
CB3 & 00 & 28 & 42.3 & +56 & 42 & 06.9 \\
CB188 & 19 & 20 & 16.1 & +11 & 35 & 57.0 \\
CB205 & 19 & 45 & 24.1 & +27 & 51 & 21.8 \\
CB230 & 21 & 17 & 39.9 & +68 & 17 & 31.9 \\
\hline
\end{tabular}
\end{flushleft}
\end{table}

\begin{figure}
\resizebox{\hsize}{!}{\rotatebox{270}{\includegraphics{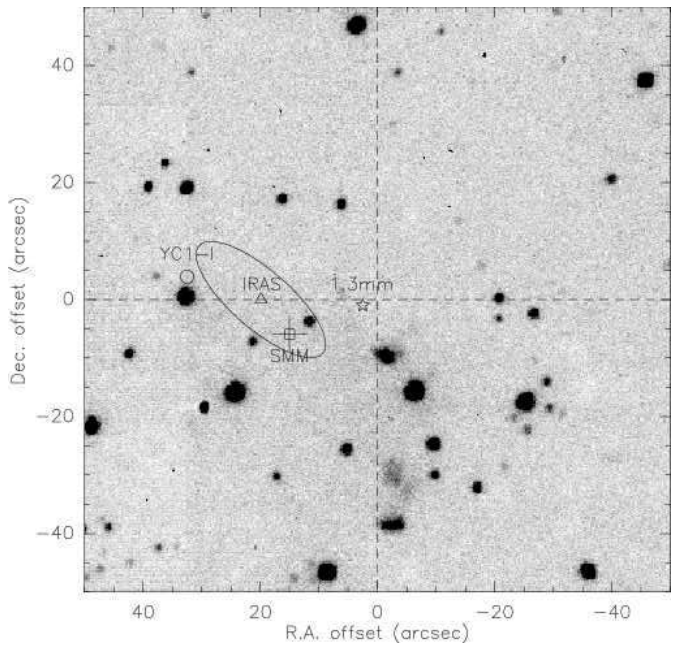}}}
\caption[]{CB3, H$_2$-band image (still including the contribution from the
continuum). We have indicated the various signposts of star formation found
associated with this globule: the NIR source YC1-I (Yun \&
Clemens~\cite{yc95};
circle), the IRAS point source (triangle), the 450~$\mu$m and 850~$\mu$m peaks
(Huard et al.~\cite{huard}; square), and the 1.3~mm source (Launhardt \&
Henning~\cite{launhen}; star). The centre of the outflow
(Codella \& Bachiller~\cite{codbach}) is at offset (0,0).
}
\label{cb3allobjects}
\end{figure}

\noindent
For each field, a set of 13 to 26 images per filter were taken on a dithered 
pattern with maximum offsets (from the centre) of 30$\arcsec$ both in 
declination and right ascension. Total on-source integration times are in the
range 20--26 min (H$_{2}$ and K$_{\rm cont}$) and 20--40 min (\FE\ and 
H$_{\rm cont}$).
Standard stars AS34-0, AS37-0 and AS40-0 (Hunt et al. \cite{hunt}) were imaged 
through all filters on a 5 position dithered pattern with the star at the 
centre of the array and within each quadrant, using integration times of 3 to 
22 s (depending on the filter and the star's brightness) per position. 
Dithered sky frames in all filters were 
acquired at dawn and sunset and used to construct differential flat fields. 
\hfill\break\noindent
Each frame was first corrected for cross-talk using the routine provided on 
the TNG web page (http://www.tng.iac.es/). Then, data reduction was done 
using standard IRAF tasks.
The obtained frames were corrected for bad pixels (through median averaging 
of the adjacent pixels) and distorsion, and then flat-fielded. For each field 
and filter, the underlying background was subtracted from each image 
in the dithered pattern by using
the median of the 6 frames nearest in time.
The same was done for the standard stars, but using 4 frames for the 
background. The images were then 
shifted to overlap, and averaged. 
Absolute coordinates were obtained through astrometry on HST Guide Stars
present in (or near) the NICS fields, as explained in Massi et al. 
(\cite{massi99}).

\section{Results and discussion}
\label{res}

\subsection{Removing the continuum}

Observations in the H$_2$ and [FeII] filters contain, apart from the emission
in the lines themselves (if present) also a contribution from the K- (in the
case of H$_2$) and H-continuum (in the case of [FeII]). 
Therefore in order to detect the line emission the continuum has to be 
subtracted. Because the central wavelength
of the continuum filters is different from those of the line 
filters, a proper subtraction requires a careful scaling of the 
emission detected in the continuum filters. Field stars are used
for this, because their emission is expected to be continuum only.

\noindent
We have assumed that the wavelength-dependence of the emission in the NIR is 
proportional to $\lambda^x$. The value of the slope $x$ is derived from a
comparison of the flux of a number of field stars 
in the line- and continuum filters. 

\noindent
In CB188 and CB205, 
no jet-like features are detected after subtraction,
whereas in CB3 and CB230 pure H$_2$ and [FeII] emission knots and
protostellar jets have been clearly revealed.
\hfill\break\noindent
Photometry of the line-emission knots has been performed on the subtracted 
images using the task {\tt POLYPHOT} in IRAF, roughly including the emission 
down to a $\sim 1 \sigma$ limit.
The sensitivity limit (of integrated line flux) is 
$2-3 \times 10^{-16}$ erg cm$^{-2}$ s$^{-1}$ arcsec$^{-1}$, both in the 
H$_{2}$ and in the [FeII] images; the limiting flux increases towards regions 
of diffuse emission.

\begin{figure*}
\resizebox{10cm}{!}{\rotatebox{270}{\includegraphics{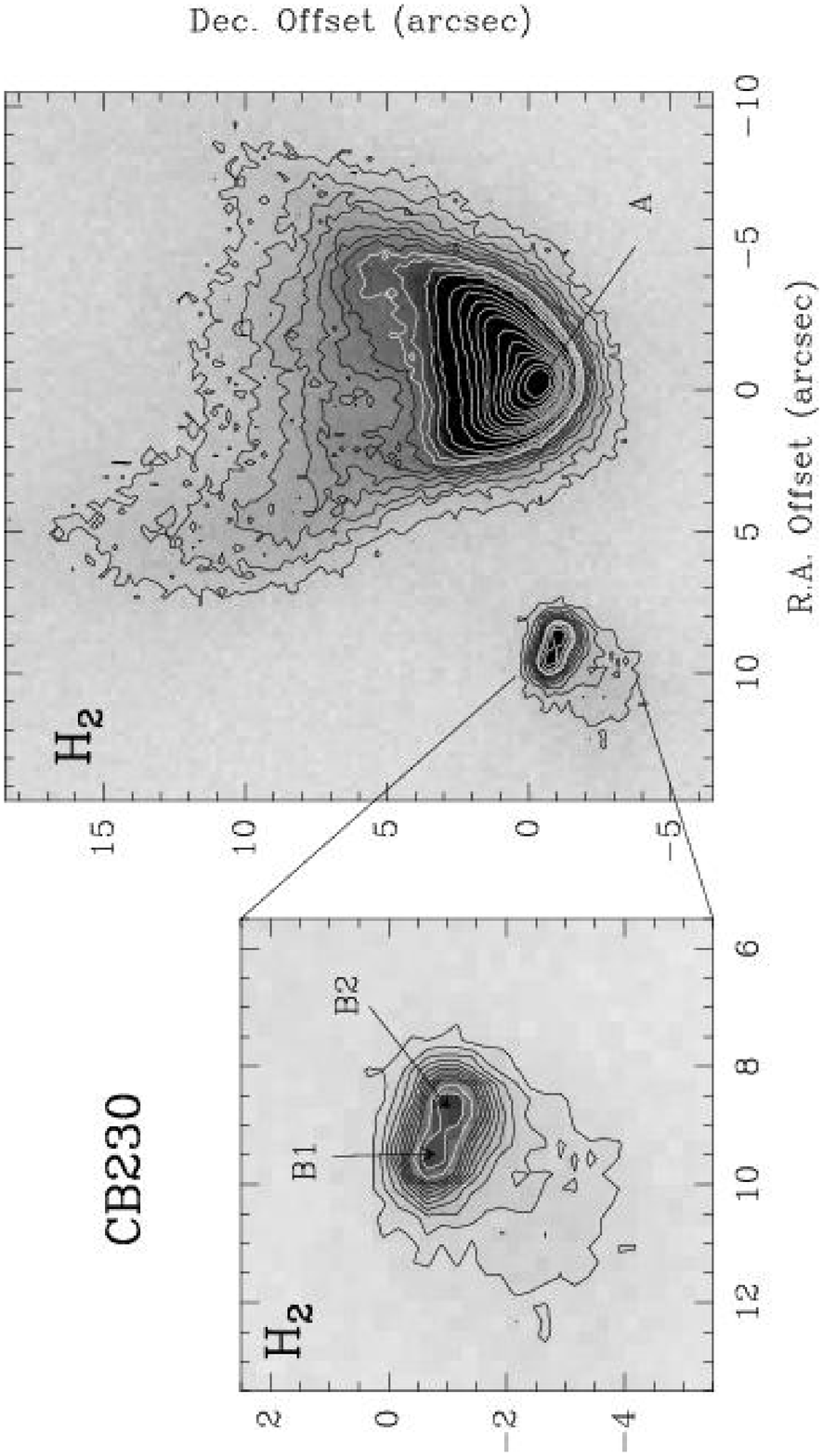}}}
\resizebox{6cm}{!}{\rotatebox{270}{\includegraphics{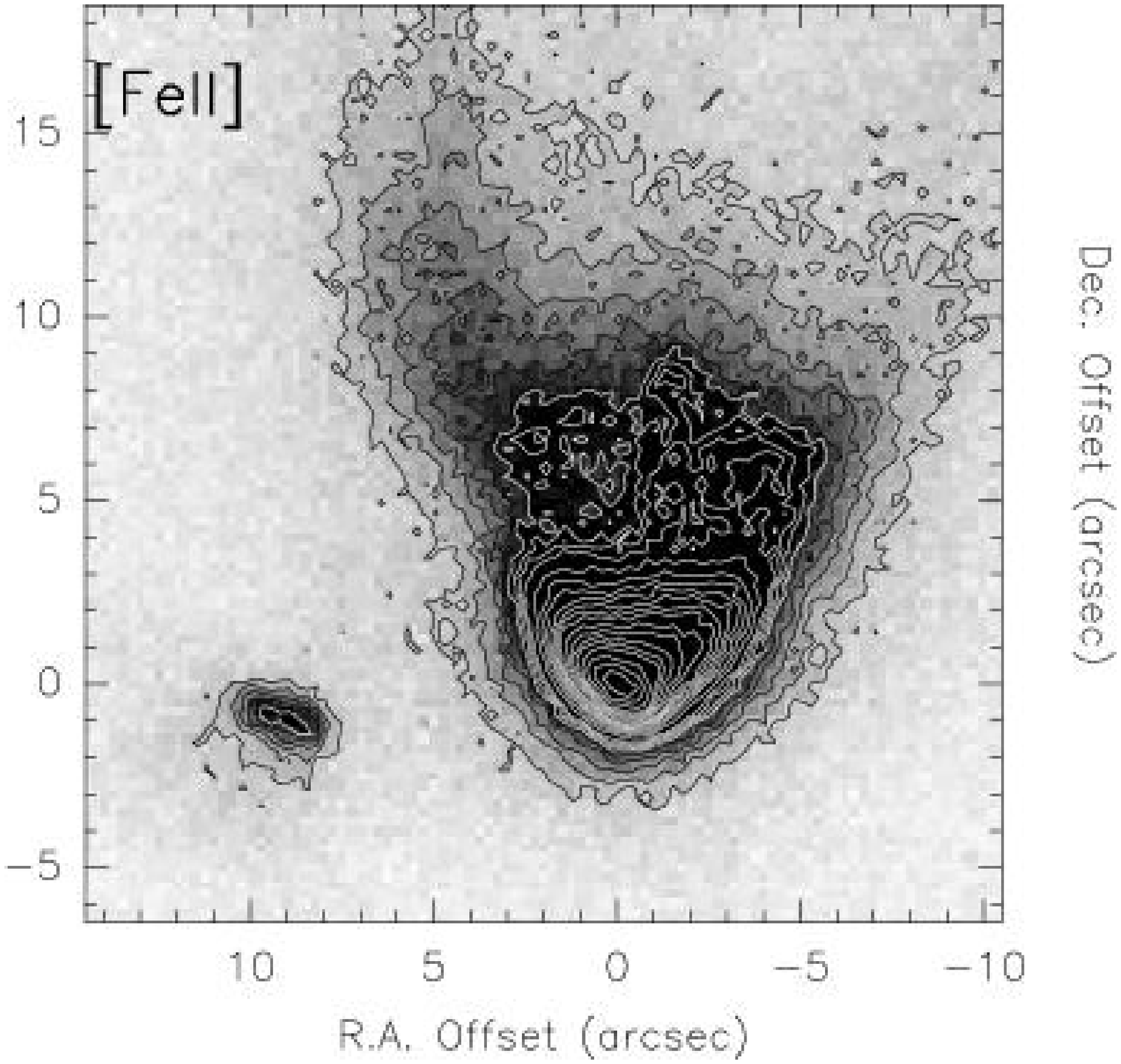}}}
\caption[]{
Results of the TNG/NICS observations of CB230 of the H$_2$ and 
[FeII]-emission (still including continuum emission):
(Central panel) A zoom-in on the
objects detected in H$_2$ in the core of CB230, with contours to better
outline the more intense emission. The YSO is indicated by ``A''. (Left)
A close-up of the smaller emission object, which is clearly seen to contain
two sources (B1 and B2). The right-hand panel shows the contoured [FeII]
emission (still including the continuum).
A clear difference is seen in the diffuse emission just North of the YSO,
with respect to the H$_2$ image (central panel); this is the signature of a
jet (see Fig.~\ref{feiih2closeup}).
}
\label{cb230h2}
\end{figure*}

\subsection{CB3}

This globule is associated with different generations of star formation, as 
pointed out by the presence of a NIR Young Stellar Object 
(YSO; Yun \& Clemens~\cite{yc95}, \cite{yc94a}),
a 1.3 mm object (Launhardt \& Henning \cite{launhen}),
and a sub-mm source (Huard et al. \cite{huard}).  
This globule also contains an IRAS point source (IRAS00259+5625), which lies 
among these objects; its flux is probably derived (in part) from contributions
by these various objects.
A molecular outflow was detected by Yun \& Clemens (\cite{yc92}, \cite{yc94b}) 
in CO and mapped in various molecular lines by Codella \& Bachiller 
(\cite{codbach}), who found that the outflow is centered at or very near the 
1.3~mm source, which is 17$\farcs$4 W, 1$\arcsec$ S of the IRAS position.

\noindent
Our NIR images are centered on the centre position of the outflow. 
Fig.~\ref{cb3h2kcont} shows the image obtained in the H$_2$  
filter. Even though this image still contains 
a contribution from the continuum, 
four regions of H$_2$ emission, hereafter called knots k1, k2, k3, and k4 
(North to South), are clearly distinguishable; they are not present in
the K$_{\rm cont}$-image (not shown). 
Fig.~\ref{cb3h2}a shows a close-up of the central region of the narrow-band 
line image; with the dashed lines we have tried to suggest how the spots can
be aligned in pairs and how they may be traced back to the centre of the 
outflow (which, within the accuracy of the positions is presumably coincident 
with the 1.3~mm source). 
The lines 
show that the centers of k1 and k3 are
aligned with the center of the outflow, while the line connecting the origin 
of the flow and the centers of k2 and k4 is slightly offset from that. This
may suggest a change of direction of the outflow axis with time. 

\noindent
Though unbeknownst to us at the time of the observations, the features k1 and 
k4 in Fig.~\ref{cb3h2}a were already
noted in deep NIR images of this region by Launhardt et al. (\cite{launISO}), 
who remarked that these two diffuse, non-stellar objects ``might have 
H$_2$-line emission contributing to their K-band luminosity''. 
The fact that these features are not visible in our K$_{\rm cont}$ image 
confirms that they arise from pure H$_2$ line emission.
The absolute coordinates of the four H$_2$ knots, as well as their
integrated flux, are given in Table~\ref{knotparams}. 

\noindent
Knot k4 lies at 
a distance of $\sim 40\arcsec$ from the centre of the outflow; for an
assumed distance of 2.5~kpc (Launhardt \& Henning~\cite{launhen}) this 
corresponds to
$\sim 0.5$~pc. Knot k1 is the nearest to the outflow centre, at a distance of
10$\arcsec \approx 0.12$~pc. All knots are found South of the outflow
centre, towards the blue lobe of the molecular outflow  
(Codella \& Bachiller \cite{codbach}). The fast outflow component is usually
traced by H$_2$ line emission, which originates behind the shocks created 
where this jet-component interacts with the ambient cloud material.
That we do not see any H$_2$ spots associated with
the red (northern) lobe of the outflow is likely because 
it is much more embedded in the globule, and the 
emission is 
too much obscured to be visible even at the NIR wavelength of H$_2$. 

\noindent
Closer inspection of the brightest knots k1 and k4 shows that the emission
contours are slightly convex. Together with the knot alignment, and the 
location of the knots along the outflow axis, this argues in favour of 
the bow shock model for the interaction of a wind from the YSO and the 
ambient medium.

\noindent
Downstream from the shocks caused by the interaction of the YSO-wind and the 
ambient medium, in high-density and -temperature gas, one expects 
to find an enhancement of several molecular species, which are liberated from 
the mantles of dust grains and injected into the gas phase through endothermic
or gas-grain reactions as well as through sputtering. 
Codella \& Bachiller (\cite{codbach}) have shown that the high-velocity 
component of the outflow in CB3 is well-traced by SiO, which is 
enhanced in the presence of shocks and which is 
only present along the main flow axis. 
In Fig.~\ref{cb3h2}b we compare the location of the H$_2$ spots with
(blue-shifted) channel maps of the emission of 
SiO(5--4). 
There is a close correspondence between the 
location of (at least two of) the H$_2$ spots and the peaks
of the SiO distribution. This confirms that in CB3 the SiO
emission is tracing the jet-like flow rather well, and indicates that along
the outflow axis different temperature regimes coexist: the warm component
at $\sim$100~K (Codella \& Bachiller 1999) traced by SiO and
a hot ($\ge$2000~K) component revealed by H$_2$.

\noindent
In Fig.~\ref{cb3allobjects} we show the locations of the various tracers of
star formation found in this globule, projected on the narrow-band H$_2$ 
image. An evolutionary sequence can be distinguished, in that we encounter
progressively younger objects when going from the 
NIR-detected object YC1 (Yun \& Clemens~\cite{yc95}), 
via the sub-mm peaks (Huard et al. \cite{huard}), 
to the 1.3~mm peak (Launhardt \& Henning~\cite{launhen}), which is probably
coincident with the centre of the outflow (Codella \& 
Bachiller~\cite{codbach}). This
implies that star formation in this globule has been going on for some
time. Assuming the IRAS source and the sub-mm peaks refer to the same
object, the mutual distance between the 3 star formation indicators in 
Fig.~\ref{cb3allobjects} is about 0.2~pc.

\noindent
We take this opportunity to point out some inconsistencies in the original
papers regarding the relative positions of the various objects plotted in 
Fig.~\ref{cb3allobjects}. First, we note that the object identified as YSO 
(based on its NIR excess) in Yun \& Clemens (\cite{yc95}) is different from
the one they identified in CB3 in Yun \& Clemens~\cite{yc94a}, which is 
about 100$\arcsec$ to the North. And even so, although YC1-I is a bright
source (K = 8.24~mag.; Yun \& Clemens~\cite{yc95}), it does not coincide with
any of the sources in our image -- the nearest object (see
Fig.~\ref{cb3allobjects}) is $\sim 4\arcsec$ to the South, much more than the
estimated positional accuracy of both our objects and YC1-I ($\sim 2\arcsec$).
\hfill\break\noindent
The relative positions of the Huard et al. 
(\cite{huard}) sub-mm peak and the IRAS source are incorrect in their Fig.~1:
based on the coordinates given by Huard et al. the sub-mm peak should be much 
closer to the IRAS position ($\sim 5\arcsec$ rather than 15$\arcsec$).
Although Launhardt \& Henning (\cite{launhen}) do not give a positional
accuracy for their 1.3~mm peak, it is probably similar to that of the sub-mm
peak (i.e. $\sim 3\arcsec$), suggesting that these are different objects.

\subsection{CB230}

This globule also hosts a single IRAS source (IRAS21169+6804), which 
is associated with the YSO found by Yun \& Clemens (\cite{yc94a}). The YSO is 
located at the apex of a cone-shaped nebulosity (see Fig.~\ref{cb230h2}) 
and is at the centre of a CO outflow (Yun \& Clemens~\cite{yc92};
\cite{yc94b}). The outflow is bipolar, but while the red component is quite
compact, the blue lobe is more elongated, and extends northwards beyond the 
boundaries of
the map made by Yun \& Clemens (\cite{yc94b}) (i.e. its length $>$3\arcmin). 
Yun \& Clemens (\cite{yc94a}) found a second NIR object, 
"possibly forming a binary system" together with the YSO at the base of
the outflow. In our observations we find this second NIR object 
to consist of two nuclei, which would make this a triple system 
(Fig.~\ref{cb230h2}). 
The conical nebula seen in 
Fig.~\ref{cb230h2} has a 
steep intensity-gradient towards the South, while towards the North the 
nebulous light fans out and decreases in intensity more gradually. The 
nebula opens up to the North, which is also the direction of the blue lobe
of the outflow, the axis of which is North-South (Yun \&
Clemens~\cite{yc94b}).
The approaching 
component of the bipolar outflow may have cleared out a cavity; the more 
extended emission seen in the northern part of the nebula could then be 
scattered light from walls of this cavity (see Yun \& Clemens~\cite{yc94a}).
Likewise, the diffuse emission associated with the embedded binary object 
$\sim 9\arcsec$ W of the main nebula is extended towards the SE (see 
Fig.~\ref{cb230h2}). This might 
be the signature of an as yet undetected (or more likely: unresolved) outflow
associated with the objects embedded herein.

\begin{figure*}
\resizebox{11cm}{!}{\rotatebox{270}{
\includegraphics{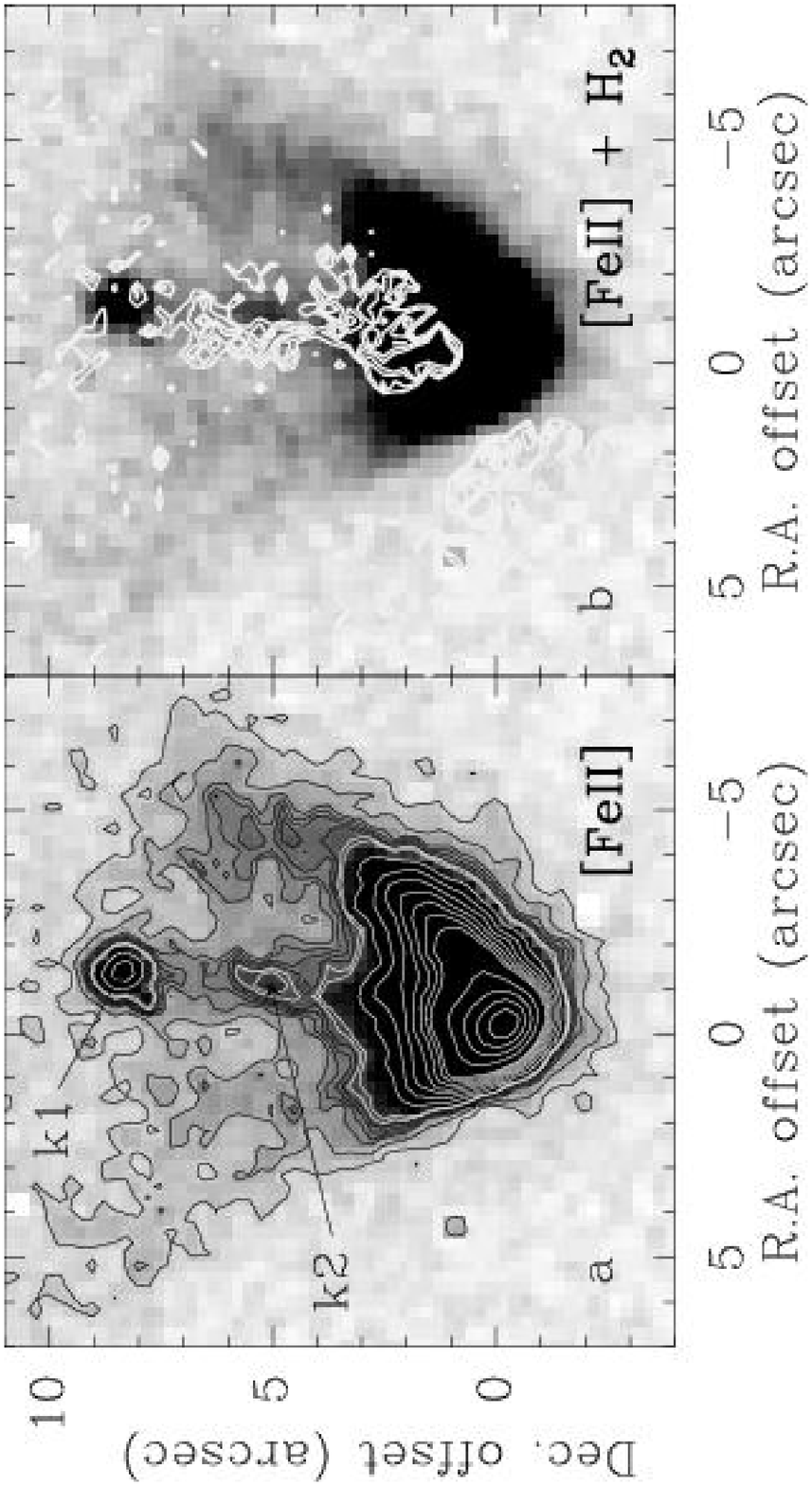}}}
\hspace{0.5cm}
\resizebox{6cm}{!}{\rotatebox{270}{
\includegraphics{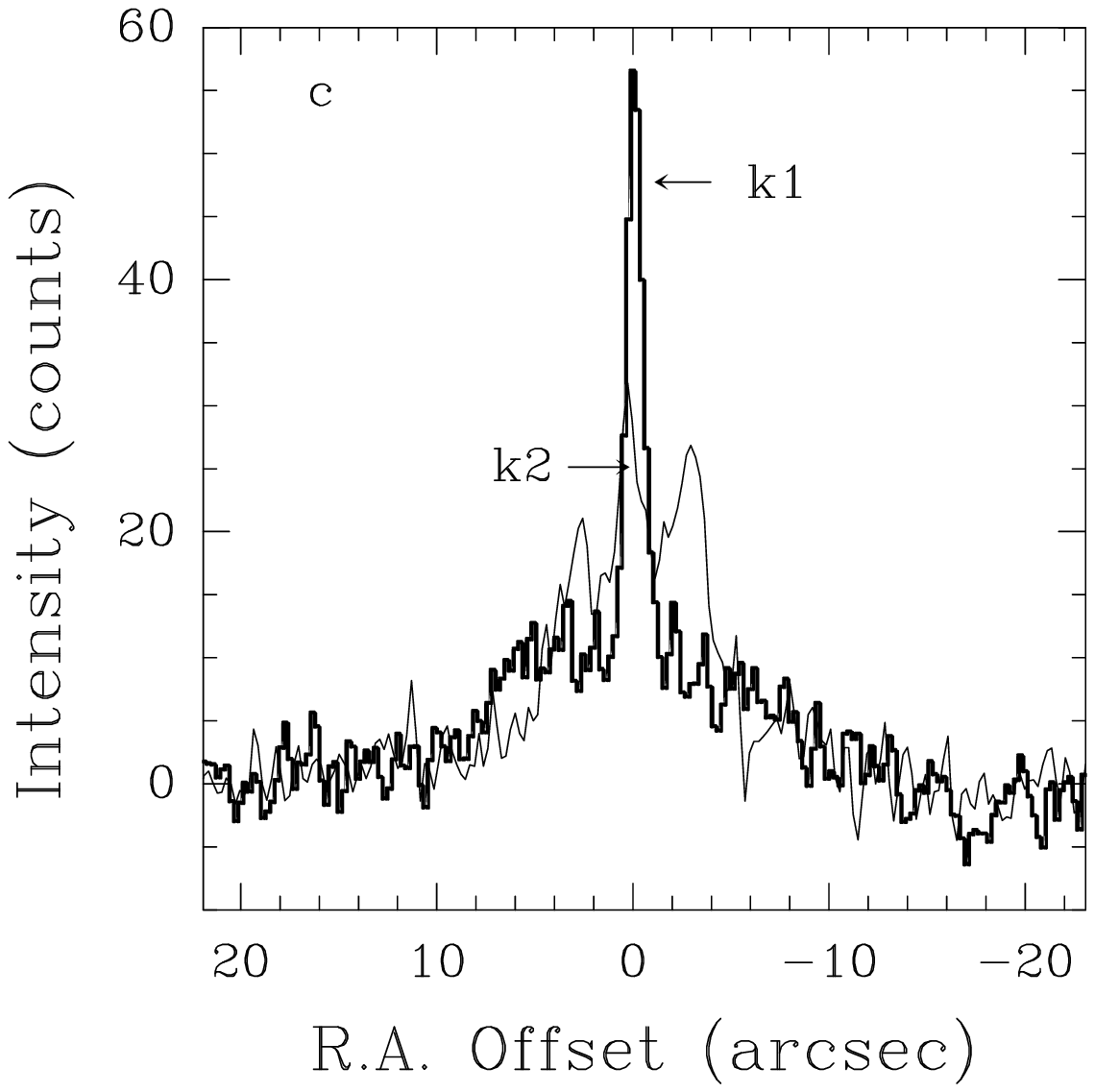}}}
\caption[]{A close-up of the jets and knots in CB230, after subtraction of
the continuum. {\bf a}\ [FeII]-emission, 
and {\bf b}\ the same but overlaid with H$_2$ contours (white).
{\bf c}\ Comparison of cuts through centre of the knots in the [FeII]-jet
at constant Dec-offset from the continuum-subtracted image. The profile for
knot k1 is drawn as a thick histogram, that for k2 as a thin line.
Most of the
elevated emission plateau between about $-8\arcsec\ {\rm and} +9\arcsec$ is an
artifact of the mosaicing together of the different frames constituting the
final image.
}
\label{feiih2closeup}
\end{figure*}

\noindent
Because we want to distinguish the (pure) line emission from the nebular 
emission, 
and because the spectral slope of the nebulosity differs from that of the
field stars, we have determined the slope $x$ of the wavelength dependence of 
the emission ($\propto \lambda^x$) by using the integrated flux of the 
northern part of the nebula, 
excluding the region with the embedded star. 
As we shall see (e.g. Fig.~\ref{feiih2closeup}) the consequence of this is 
that in the continuum-subtracted [FeII]-image the YSO is still visible.
Its spectral index differs from that of the nebula (which is 
radiation scattered by the dust), which we have used to scale the 
H$_{\rm cont}$-image; the YSO cannot therefore be correctly subtracted.

\noindent
The final reduction, with careful scaling of the H$_{\rm cont}$- and 
K$_{\rm cont}$ images of
the diffuse emission, reveals the pure line emission that remains after
subtraction of the continuum, and is shown in Fig.~\ref{feiih2closeup}: a 
jet is seen in [FeII], primarily defined by two knots along its
length: a bright one at the tip (hereafter called k1) and a fainter one (k2)
about half-way between k1 and the YSO. The knots are superimposed on a
fainter, but clearly visible elongated emission feature.
We stress that this detection is independent of
potential inaccuracies in the continuum subtraction, as the knots are visible
in the contours of the [FeII] image before continuum subtraction (cf. 
Fig.~\ref{cb230h2}), but not at all in the (H$_{\rm cont}$) continuum image 
alone. 
\hfill\break\noindent
The jet is oriented in the N-S
direction, and lies at the base of the large-scale molecular outflow.
This jet is also traced by H$_2$ emission (Fig.~\ref{feiih2closeup}b), which 
is weaker and slightly less narrow than that of [FeII]. Moreover, the 
H$_2$ emission appears to be anti-correlated with [FeII] peaking 
next to, rather than on the secondary knot in [FeII], and
we conclude that in H$_2$ we probably see the {\sl walls} of the
jet-channel. The detection of strong [FeII]- and weaker H$_2$
emission suggests the presence of fast, dissociative J-shocks.

\noindent

The total length of the jet, from 
its base at the location of the embedded YSO to the edge of knot k1 is about
9\pas5, corresponding to $\sim 0.02$~pc at the assumed distance of 450~pc 
(Launhardt \& Henning~\cite{launhen}).
Profiles of the knots in the [FeII]-jet, along cuts made at constant 
declination offsets and passing through the location of peak intensity of the 
knots, are shown in Fig.~\ref{feiih2closeup}c.
Both features are
resolved in the NIR image.
The width (FWHM) of k1 is 1\pas2\, corresponding to $\sim 540$~AU. In the
profile of k2 one can also identify the two intensity enhancements to the
East and West, seen in Fig.~\ref{feiih2closeup}a, which deliniate the edges
of the conical nebula. Both knots are superimposed
on a broad ($\sim 17\arcsec \approx 0.04$~pc) plateau of emission; this is
an artifact due to imperfect mosaicing, rendering the middle section of the
image (in which the nebula and the jet are located) more noisy than the edges. 
The absolute coordinates of the three stellar objects 
(A, B1, and B2; Fig.~\ref{cb230h2}) and of the two knots 
(k1, k2; Fig.~\ref{feiih2closeup}a) in the [FeII]-jet are given in 
table~\ref{knotparams}, where we also list the integrated fluxes of the two 
[FeII]-knots and the entire H$_2$ jet.
\hfill\break\noindent
In addition to the main jet, a second [FeII]-feature is seen, emanating from
the bright nebulosity and oriented in a N-NW-direction; this may indicate the
existence of a second (unresolved) outflow, or it 
is a shock at the surface of the cavity, and thus related to the main jet.

\begin{table}
\caption[] {List of coordinates and fluxes of the detected objects}
\label{knotparams}
\begin{tabular}{lcccc}
\hline
\multicolumn{1}{c}{Object} &
\multicolumn{1}{c}{$\alpha_{\rm 2000}$} &
\multicolumn{1}{c}{$\delta_{\rm 2000}$} &
\multicolumn{1}{c}{$F_{\rm [FeII]}$} &
\multicolumn{1}{c}{$F_{\rm H_2}$} \\ 
\multicolumn{1}{c}{CB} &
\multicolumn{1}{c}{($^h$ $^m$ $^s$)} &
\multicolumn{1}{c}{($\degr$ $\arcmin$ $\arcsec$)} &
\multicolumn{2}{c}{(erg cm$^{-2}$ s$^{-1}$)$^{\dagger}$} \\
\hline
3-k1  &  00 28 42.17 & +56 41 57.36 & $<$ 2.4$-$16 & 3.5$-$14 \\ 
3-k2  &  00 28 42.12 & +56 41 37.17 & $<$ 2.4$-$16 & 1.4$-$14 \\
3-k3  &  00 28 41.75 & +56 41 34.20 & $<$ 2.4$-$16 & 5.6$-$15 \\
3-k4  &  00 28 42.12 & +56 41 28.56 & $<$ 2.4$-$16 & 2.1$-$14 \\
205-A  & 19 45 24.15 & +27 50 57.18 & -- & --  \\
205-B  & 19 45 23.99 & +27 50 59.25 & -- & --  \\
230-A  &  21 17 38.36 & +68 17 32.87 &  -- & --  \\ 
230-B1 &  21 17 40.09 & +68 17 32.15 &  -- & --  \\
230-B2 &  21 17 39.96 & +68 17 31.91 &  -- & --  \\
230-H$_2$ &     --       &    --        &  --  & 1.1$-$14 \\ 
230-k1 &  21 17 38.11 & +68 17 41.04 & 6.4$-$15 &  --  \\
230-k2 &  21 17 38.12 & +68 17 38.08 & 3.2$-$15 &  --  \\
\hline
\multicolumn{5}{l}{$\dagger$ a$-$b means a$\times 10^{-b}$}
\end{tabular}
\begin{center}
\end{center}
\end{table}

\subsection{CB188}

\begin{figure*}
\resizebox{15cm}{!}{\rotatebox{270}{
\includegraphics{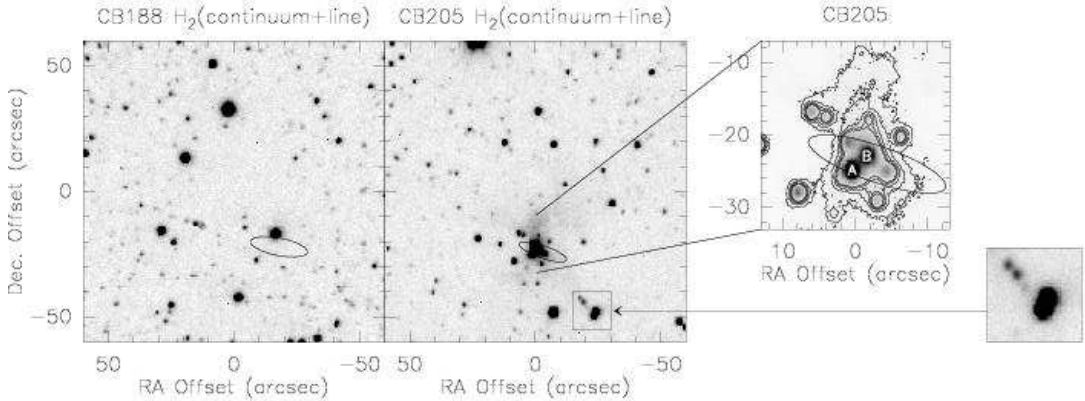}}}
\caption[]{Results of the TNG/NICS observations showing the H$_2$ 
emission, still including continuum emission, towards CB188 (left panel) and 
CB205 (right panel). The locations of the IRAS point sources are indicated by 
their uncertainty ellipses. The upper small panel on the far right shows a 
blow-up of the region around the IRAS source position in CB205.
The lower small panel on the far right shows the alleged jet-like
feature (Yun et al.~\cite{yetal93}) being resolved into stars.
}
\label{cb188205}
\end{figure*}

In this globule, Launhardt \& Henning (\cite{launhen}) detected 1.3~mm
continuum emission, in the form of an ``extended source with compact
components'' at the location of the IRAS source (indicated by the ellipse in
Fig.~\ref{cb188205}). A YSO was detected at the same location by Yun \&
Clemens (\cite{yc95}), while Yun \& Clemens (\cite{yc94b}) mapped a small 
($\leq 2\arcmin$) outflow, of which the less extended blue lobe overlaps the 
red one. 

\noindent
Our continuum-subtracted H$_2$ and [FeII]-images do not 
show any line emission in the field down to the sensitivity limit.
We also do not see the cometary diffuse emission around the
possible NIR counterpart of the IRAS source that was found by
Yun \& Clemens (\cite{yc94a}; this would be the bright star seen just
outside the IRAS error ellipse in Fig.~\ref{cb188205}), nor do we detect their 
source F, which they find embedded in this diffuse emission (see their 
Fig.~10b). It might be either a spurious detection or a variable object.

\subsection{CB205}

This globule, also known as L810, has an extended NIR nebulosity at the 
location of the IRAS source, as seen in Fig.~\ref{cb188205}. A NIR star
cluster is associated with it; at least 10 stars can be seen within 
$10\arcsec$ of the IRAS position (see the top-inset of Fig.~\ref{cb188205}). 
\hfill\break\noindent
The major axis of the nebula is oriented in a N-S direction, which is also
the orientation of the compact molecular outflow detected by Yun \& Clemens
(\cite{yc94b}), of which the red and blue lobes show a significant overlap.
This coincidence of orientations suggests a connection between the shape
of the nebula and the presence of the outflow. However, no line emission has
been found in our continuum-subtracted images, although we cannot rule out
that some line emission knots may have been hidden in the more intense parts
of the diffuse NIR nebulosity, near the IRAS point source location.

\noindent
The nebula has been studied in the NIR (J,H,K) by Yun et al. (\cite{yetal93}).
The object they identified as the illuminator of the nebula, L810IRS, 
coincides with our star B in Fig.~\ref{cb188205}. Yun et al.
(\cite{yetal93}) also 
report the detection of an elongated jet-like feature, located about 
$35\arcsec$ from L810IRS, to the SW. This feature is also visible in all our
images (both line and continuum) and is seen to be resolved in a coincidental 
alignment of stars embedded in diffuse emission (see the lower-inset in 
Fig.~\ref{cb188205}, at 
$\Delta\alpha \sim -20\arcsec$ and $\Delta\delta \sim -45\arcsec$).
No emission is left after continuum subtraction, further excluding
a jet origin for it.

\section{Summary}
We have used NICS at the TNG to search for jet-components associated with 
molecular outflows in 
four globules, by imaging through narrow-band filters centered at [FeII] and
H$_2$ and the adjacent continuum. Jets can be revealed, if present, after 
subtracting the scaled continuum emission from the image taken through the
line filters. In two of our targets, CB3 and CB230, this led to the 
identification of regions of pure line emission, while nothing was found in
CB188 and CB205.

\noindent
In CB3 we found four regions (``knots'') of H$_2$ emission. All of these
are located in the southern, blue lobe of the molecular outflow, which is
probably driven by the 1.3~mm source located very near the center of the 
outflow. There is a close correspondence between the location of the 
H$_2$-knots and the peaks in the SiO(5$-$4) emission 
(Codella \& Bachiller~\cite{codbach}). 
The knots likely identify the locations where the 
jet-component of the outflow hits and shocks the ambient gas. 
\hfill\break\noindent
The relative locations of various signposts of star formation (embedded NIR
source, IRAS point source, sub-mm and mm-peaks, and the outflow) suggest that 
star formation has been active for some time in this globule, and is 
progressing from East to West in CB3.

\noindent
In CB230 we find a jet emerging from a conical nebulosity in which a YSO is
embedded. The jet is oriented N-S, in the same direction as the large-scale 
molecular outflow. In the [FeII]-line the jet consists of two knots, the 
brighter one of which
lies at its tip. Its length from the YSO at its base to its tip is about 
0.02~pc (for $d=450$~pc). The knots are superimposed on a substrate of 
lower-level diffuse emission. H$_2$ line emission is also detected, but this 
is fainter, and slightly wider than that of [FeII], suggesting that here
we see the walls of the cavity in which the jet itself flows.
\hfill\break\noindent
A second, fainter [FeII]-feature is seen to emerge from the nebulosity, and
pointing in a NW-direction. This might either indicate the presence of a
second, unresolved outflow, or be the locus of a shock at the surface of 
the cavity blown by the outflow.
\hfill\break\noindent
About $9\arcsec$ East of the YSO/nebula/jet-complex we find a smaller 
nebulosity containing two star-like nuclei. The nebulosity in which these are
embedded has a steep intensity gradient to the NW, and shows a more gentle 
decrease towards the SE, analogous to what is seen in the main nebula. 
From the existing low-angular resolution CO maps it is not possible to tell
if this binary object is associated with its own outflow. 

\begin{acknowledgements}
The work presented here is 
based on observations made with the Italian Telescopio Nazionale Galileo
(TNG) operated on the island of La Palma by the Centro Galileo Galilei of
the INAF (Istituto Nazionale di Astrofisica) at the Spanish Observatorio del
Roque de los Muchachos of the Instituto de Astrofisica de Canarias. A 
special thank you to Dino Fugazza and Gianni Tessicini 
for their help during the observations. We thank Rafael Bachiller for
suggestions and stimulation discussions.
\end{acknowledgements}


\begin{thebibliography}{}

\bibitem[1993]{allen}
Allen D.A., Burton M.G., 1993, Nature 363, 54
\bibitem[1996]{bach}
Bachiller R., 1996, ARA\&A 34, 111
\bibitem[2001]{baffa}
Baffa C., Comoretto G., Gennari S., et al., 2001, A\&A 378, 722
\bibitem[1988]{clemens}
Clemens D.P., Barvainis R.E., 1988, ApJS 68, 257 [CB]
\bibitem[1999]{codbach}
Codella C., Bachiller R., 1999, A\&A 350, 659 
\bibitem[2000]{huard}
Huard T.L., Weintraub D.A., Sandell G., 2000, A\&A 362, 635
\bibitem[1998]{hunt}
Hunt L.K., Mannucci F., Testi L., et al., 1998, ApJ 115, 2594 
\bibitem[1997]{launhen}
Launhardt R., Henning Th., 1997, A\&A 326, 329
\bibitem[1998]{launISO}
Launhardt R., Henning Th., Klein R., 1998, in ``Star formation with the 
Infrared Space Observatory (ISO)'' (eds. J.L. Yun, R. Liseau), ASP Conf. Ser. 
132, p.119
\bibitem[1999]{massi99}
Massi F., Giannini T., Lorenzetti D., et al., 1999, A\&AS 136, 471
\bibitem[1992]{yc92}
Yun J.L., Clemens D.P., 1992, ApJ 385, L21
\bibitem[1994a]{yc94a}
Yun J.L., Clemens D.P., 1994a, AJ 108, 612
\bibitem[1994b]{yc94b}
Yun J.L., Clemens D.P., 1994b, ApJS 92, 145
\bibitem[1995]{yc95}
Yun J.L., Clemens D.P., 1995, AJ 109, 742
\bibitem[1993]{yetal93}
Yun J.L., Clemens D.P., McCaughrean M.J., Rieke M., 1993, ApJ 408, L101
\end{thebibliography}
\end{document}